# Towards a Centralized Scheduling Framework for Communication Flows in Distributed Systems

**Mugurel Ionut Andreica\*, Eliana-Dina Tirsa\*, Nicolae Tapus\*, Florin Pop\*, Ciprian Mihai Dobre\***

\* Computer Science and Engineering Department, Politehnica University of Bucharest, Bucharest, Romania
(e-mail: {mugurel.andreica, eliana.tirsa, nicolae.tapus, florin.pop, ciprian.dobre}@cs.pub.ro)

**Abstract:** The overall performance of a distributed system is highly dependent on the communication efficiency of the system. Although network resources (links, bandwidth) are becoming increasingly more available, the communication performance of data transfers involving large volumes of data does not necessarily improve at the same rate. This is due to the inefficient usage of the available network resources. A solution to this problem consists of data transfer scheduling techniques, which manage and allocate the network resources in an efficient manner. In this paper we present several online and offline data transfer optimization techniques, in the context of a centrally controlled distributed system.

## 1. INTRODUCTION

Large distributed systems in which significant volumes of data are routinely transferred are becoming more frequently deployed and more prevalent nowadays. The communication performance of such systems has a strong impact upon their overall efficiency and, thus, a great emphasis is placed on the development of efficient communication optimization techniques. In this paper we consider several online and offline data transfer and data distribution optimization problems, in the context of a centrally controlled distributed system. Although recent research results in this field focus on large scale decentralized systems, we argue that, in real life situations, these systems are actually composed of multiple centrally controlled distributed systems and, thus, focusing on systems with centralized control is a matter of practical interest. The online data transfer optimization problems are considered in the context of a centralized data transfer scheduling framework which is introduced in Section 2. However, the focus of this paper is not on the actual scheduling framework, but on the algorithmic techniques which are employed by its central component, the *Communication Flow Scheduling and Optimization Component*. In Sections 3 and 4 we present algorithmic results regarding data transfer scheduling on single network links and in tree networks. In Section 5 we present efficient algorithmic solutions for some offline data distribution problems. In Section 6 we present related work and in Section 7 we conclude and discuss future work.

## 2. DATA TRANSFER SCHEDULING FRAMEWORK

The online data transfer scheduling model which is considered by the scheduling framework introduced in this section was previously described in (Andreica and Tirsa, 2008) and (Andreica and Tapus, 2008). A centralized scheduler has full control over all (or most of) the traffic in the network. Each network node can submit data transfer requests to the scheduler. A request may contain several parameters, like: *source node*, *destination node(s)*, *start time*, *finish time*, *duration*, *minimum required bandwidth* (in the case of non-preemptive data transfers), *total size of the transferred data* (in the case of preemptive data transfers), *profit* (obtained if the request is scheduled and all of its constraints are satisfied). The scheduler handles the requests in batches of at most $R \geq 1$ requests at a time (the scheduler waits until the number $m$ of received requests equals $R$ or until a short time limit is exceeded, if $1 \leq m < R$). Once a batch of requests is constructed, the scheduler runs an optimization algorithm, considering the $m \leq R$ requests in the batch, as well as the previously scheduled data transfers. The scheduler may consider that time is divided into $T$ equally-sized time slots ($T$ is the time horizon over which data transfers can be scheduled) or may consider only the time moments when an event occurs (e.g. a data transfer starts or ends).

The scheduler is only a component of a data transfer scheduling framework which needs to be developed in order to seamlessly provide data transfer scheduling and optimization services. The framework consists of several components: (1) the *Communication Flow Scheduling and*

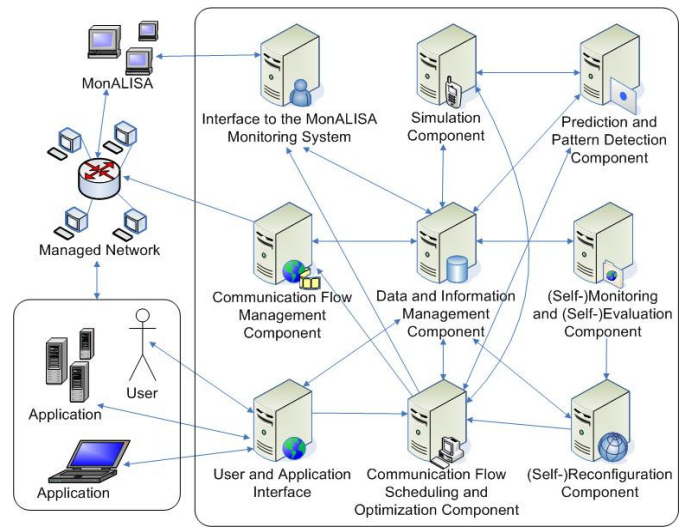

Fig. 1. Architecture of the Data Transfer Scheduling Framework

Optimization Component ; (2) the Data and Information Management Component ; (3) the Communication Flow Management Component ; (4) the User and Application Interface ; (5) Interface to a Monitoring System (e.g. MonALISA) ; (6) the Prediction and Pattern Detection Component ; (7) the Simulation Component ; (8) the (Self-) Monitoring and (Self-) Evaluation Component ; (9) the (Self-) Reconfiguration Component. Fig. 1 presents all the components, together with the directions of the command and data flows between them. We intend to use the MonALISA monitoring system (Legrand et al., 2004) to provide monitoring data to the scheduling framework (i.e. information about the relevant network parameters and about the status of the running data transfers). The core of the framework is the Communication Flow Scheduling and Optimization Component, which runs the optimization algorithms and makes the scheduling decisions. This component may use simulations (the Simulation Component) or pattern detection and data transfer request prediction techniques (the Prediction and Pattern Detection Component) in order to make improved scheduling decisions. The decisions of this component are transformed into commands for the network nodes by the Communication Flow Management Component. The Data and Information Management Component stores all the data of the framework and, as such, it is connected to all the other components. The (Self-) Monitoring and (Self-) Evaluation Component monitors the quality of the decisions made by the scheduling component. If they are not of sufficient quality, it may use the services of the (Self-) Reconfiguration Component in order to reconfigure the Communication Flow Scheduling and Optimization Component (e.g. change the scheduling algorithm, switch from a time-slot based to an event-based time interpretation). At this point, the proposed framework is only in prototype stage. The focus of the rest of this paper is on algorithmic techniques, some of which can be implemented and used by the Communication Flow Scheduling and Optimization Component.

## 3. DATA TRANSFERS ON A SINGLE NETWORK LINK

In this section we will consider that the scheduler uses the time slot-based model and maintains a value $avb(t)$ for each time slot $t$ ($1 \leq t \leq T$), representing the available bandwidth within that time slot (initially, all the values are equal to $B_{max}$, the maximum bandwidth of the network link). A data transfer request $r$ consists of the following parameters: starting time slot ($S(r)$), finish time slot ($F(r)$), total amount of data to be transmitted ($TD(r)$) (if it is preemptive) or the minimum required bandwidth $B$ (if it is non-preemptive). In the preemptive case, a request is granted if we can assign a bandwidth $b(r,t)$ to every time slot $S(r) \leq t \leq F(r)$, such that: the sum $b(r,S(r)) + b(r,S(r)+1) + \ldots + b(r,F(r)) = TD(r) / slot\_d$ ($slot\_d$=the duration of a time slot) and $b(r,t) \leq avb(t)$, for $S(r) \leq t \leq F(r)$. If the request is granted, the values $avb(t)$ are decreased by $b(r,t)$ for every time slot $t$ in the range. For the non-preemptive case, we consider two types of requests. The first type requires us to assign the bandwidth $b(r,t)=B$ during every time slot $t$ ($S(r) \leq t \leq F(r)$) (and the transfer duration is $F(r)-S(r)+1$). The second type has unit duration and asks us to find only one time slot $t$ ($S(r) \leq t \leq F(r)$) where we can assign $b(r,t)=B$. We will consider two scheduling models: the *batch* model (where multiple requests are considered at a time) and the *online* model (where we consider one request at a time).

### 3.1 Preemptive Data Transfer Requests in Batches

The $m \leq R$ requests of a batch are handled as a group. Each request $r$ also has a profit $p(r)$, representing the profit gained if the request is granted. We propose here a heuristic method which tries to maximize the total profit of the accepted requests from each group. We will construct a bipartite graph containing the $m$ requests on the left side and the $T$ time slots on the right side. We will add an edge with infinite capacity from every request $r$ to every time slot $t$ in the time slot interval $[S(r), F(r)]$. We will now add two extra vertices, *src* and *dest*. We add an edge from *src* to every request $r$ and assign to it a capacity equal to $TD(r)/slot\_d$. We then add an edge from every time slot $t$ to the vertex *dest* and assign to it a capacity equal to $avb(t)$. We will now find the maximum flow $F$ from *src* to *dest* in this graph. If the flow $f(src,r)$ on every edge $(src,r)$ is equal to the edge's capacity, then all the requests can be granted. In this case, the flow $f(r,t)$ on every edge between a request $r$ and a time slot $t$ represents the value $b(r,t)$. If we cannot accept all the requests in the batch, we have two options. The first one consists of granting the requests $r$ with $f(src,r)=TD(r)/slot\_d$ (if any) and re-run the algorithm for the remaining requests (considering the updated values $avb(t)$) repeatedly, until no more requests can be granted. For the second option we will define a desirability function $d(r)$ which considers the profit of the requests, the amount of data to transfer, the transfer duration and possibly other parameters. For instance, such a function could be $d(r)=p(r) \cdot (F(r)-S(r)+1)^{exp}/TD(r)$ ($exp>0$). We will sort the requests in non-increasing order of their desirability: $r_1, r_2, \ldots, r_m$, such that $d(r_1) \geq d(r_2) \geq \ldots \geq d(r_m)$. We will find the index $1 \leq q \leq m+1$, s. t. if the batch consisted of the set of requests $\{r_1, \ldots, r_{q-1}\}$, then all of them could be accepted, but if the batch consisted of the set $\{r_1, \ldots, r_q\}$, then not all of them could be accepted. We can find $q$ by using binary search (or linear search) and the maximum flow algorithm presented above (in order to decide if all the request in a set $\{r_1, \ldots, r_p\}$ can be accepted and, consequently, decide if $q>p$ or $q \leq p$). Then, we will accept the requests $r_1, \ldots, r_{q-1}$, reject the request $r_q$ and re-run this algorithm for a batch composed of the remaining requests $r_{q+1}, \ldots, r_m$ (considering the updated $avb(t)$ values).

### 3.2 Online Non-Preemptive Data Transfer Requests

For the first type of non-preemptive requests, we need to find the minimum available bandwidth within the time slot interval $[S(r),F(r)]$. For this purpose we can use the segment tree framework or the block partitioning framework, presented in (Andreica and Tapus, 2008). Both frameworks support range minimum queries (computing the minimum available bandwidth in the interval $[S(r),F(r)]$, in order to compare it against the minimum required bandwidth) and range addition updates (decreasing the available bandwidth of all the time slots in a range $[S(r),F(r)]$ by the same value $B$) in sublinear time ($O(log(T))$ for the segment tree and $O(k+T/k)$ for the partition into blocks). Note that (Andreica and Tapus, 2008) incorrectly claims that these types of operations (range addition update and range minimum query) can be used in the case $ES=1, LF=m$; instead, they work only for $LF-ES+1=D$ (using their notation). For the second type of non-preemptive requests (having unit durations), we could use again the frameworks from (Andreica and Tapus, 2008), with the range (or point) addition update and range maximum query operations. It is obvious that the maximum available bandwidth of a slot inside the interval *[S(r),F(r)]* is the best „answer" for a query. However, always choosing the slot with the maximum available bandwidth may cause future requests to be rejected because not enough bandwidth is available. In these cases, we might want to assign to a request asking for a minimum bandwidth *B* a time slot with just enough available bandwidth, leaving the slots with large amounts of available bandwidth to requests with large bandwidth requirements. In order to implement this behaviour, we split the *T* time slots into *T/k* groups of *k* time slots each (the last group may contain less than *k* time slots). For each group *G*, we will maintain all the available bandwidths of its time slots sorted in ascending order. For a range of time slots *[S(r),F(r)]*, we classify every group *G* as:

- G is completely inside *[S(r),F(r)]* – internal group
- G is completely outside *[S(r),F(r)]* – external group
- G is partly crossing *[S(r),F(r)]* – partially crossing group

There can be at most two partially crossing groups (at the left and right sides of the range *[S(r),F(r)]*). In the case of an (addition) update, the time slots of each partially crossing group *G* which are inside *[S(r),F(r)]* are modified and then all the time slots inside *G* are resorted. For every internal group *G*, we will add the update value to the *globalbw* field of *G* (this field is initially *0*). When searching for a value larger than *B* inside a range *[S(r),F(r)]*, we will test the real value of every time slot *t* (*avb(t)+G.globalbw*) inside the intersection of a partially crossing group *G* and *[S(r),F(r)]*; for every internal group *G*, we will binary search within the values of its time slots (which are sorted) the smallest value larger than *(B-G.globalbw)*. Every update takes $O(T/k+k \cdot log(k))$ time and every query takes $O(k+T/k \cdot log(k))$ time. If an exact value *B* is searched, we can maintain a hash table with the available bandwidths of the time slots in every group (instead of a sorted array) and replace the binary search by a hash lookup and the resorting process by a hash rebuilding; thus, both updates and queries would take $O(k+T/k)$. In the case of the first type of non-preemptive data transfer requests (actually, a slightly more general case), a special subcase was considered in (Andreica and Tapus, 2008). This subcase occurs when every request asks for the whole bandwidth of the link (thus, concurrent data transfers cannot take place). A simple solution based on maintaining a balanced tree of empty and occupied time slot intervals was proposed there. An update is equivalent to coloring an entire interval *[a,b]* with the same color (*1*-if a reservation is placed; *0*-if a reservation is cancelled). The update procedure had a minor flaw there, which we correct in this section. The balanced tree *BT* contains a set of maximally-colored disjoint intervals whose union is *[1,T]*. Initially, *BT* contains only one interval, *[1,T]*, colored with *0*. When coloring an interval *[a,b]* with a color *col*, we first find all the intervals *[c,d]* in *BT* which are fully included in *[a,b]* and we remove them from *BT*. Then, if *[a,b]* is fully included in an interval *[c,d]* in *BT* with color *col'*, we remove *[c,d]* from *BT* and insert in *BT* the intervals *[c,a-1]* (if $c \leq a-1$) and *[b+1,d]* (if $b+1 \leq d$), colored with *col'*. Then, we find the (at most) two intervals *[p(j),q(j)]* (*j=1,2*) which partially intersect *[a,b]* (at the left endpoint and at the right endpoint) (i.e. ($p(j)<a$ and $a \leq q(j)<b$) or ($a<p(j) \leq b$ and $q(j)>b$)). Let *[p'(j),q'(j)]* be the intersection of *[p(j),q(j)]* with *[a,b]*. We remove *[p(j),q(j)]* from *BT* and insert in *BT* the interval *[p(j),q(j)]\[p'(j),q'(j)]* (the part of *[p(j),q(j)]* which does not intersect *[a,b]*), colored with the same color as *[p(j),q(j)]*. Then, we insert *[a,b]* in the tree. The final step, which was forgotten in (Andreica and Tapus, 2008) is to find the interval *[c,a-1]* in *BT*, located immediately to the left of *[a,b]* (if it exists) and check if it has the same color as *[a,b]*. If it does, then we remove the intervals *[c,a-1]* and *[a,b]* from *BT* and insert in *BT* the interval *[c,b]* (with the same color *col*); if the replacement was performed, we set *a=c*. Then, we check if the interval *[b+1,d]* from *BT*, located immediately to the right of *[a,b]* (if it exists), has the same color as *[a,b]*; if it does, we remove both intervals from *BT* (*[a,b]* and *[b+1,d]*) and insert in *BT* the interval *[a,d]* (having the same color as the two intervals). This final step ensures that the intervals in *BT* are maximally-colored (i.e. *BT* does not contain two adjacent intervals with the same color). The interval coloring problem has several other variants, like the following. We are given *M* coloring operations which must be performed sequentially: color the interval of time slots *[a(i),b(i)]* with color *col(i)* (*col(i)* is not necessarily *0* or *1*). After performing all the operations, we need to find the final color of each time slot. Initially, all the time slots have the color *colinit* (e.g. *colinit=0*). We could use the same balanced tree *BT* presented before, which maintains maximally-colored intervals. After every coloring operation, the number of intervals in *BT* increases by at most *2*. Thus, the overall time complexity is $O(M \cdot log(M))$. Since we know all the coloring operations in advance, we can use another technique. We sort all the left and right endpoints of the coloring intervals in increasing order (if a left and a right endpoint have the same value, then we place the left endpoint before the right endpoint) and assign to each operation the value *k*, if it is the $k^{th}$ operation in the sequence. We also consider the interval *[1,T]*, with value *0*. We then traverse the endpoints of the intervals in the sorted order. When we reach the left endpoint (slot) *t* of an interval *j*, we will insert *j* into a max-heap *H*; the keys of the intervals in *H* are their values (i.e. their positions in the sequence of coloring operations); *H* will always contain a "fake" interval with -∞ value and any color. After processing all the left endpoints equal to *t*, we find the interval *i* in *H* with the largest value *v(i)* and produce the tuple *(t, v(i), col(i))* (*col(i)* is the color of the operation corresponding to the interval *i*). When we reach the right endpoint *t* of an interval *j*, we remove the interval *j* from *H*; after processing all the right endpoints equal to *t*, we produce the tuple *(t+1, v(i), col(i))* (*i* is the interval with the largest value *v(i)* in *H*). Afterwards, we consider all the tuples *(t(j), value(j), color(j))*, in the order in which they were produced. If multiple tuples have the same *t* field, we will keep only the tuple with the largest *value* field among them and remove the others. Let's consider these tuples in the order *(t(1), value(1),*

*color(1)), ..., (t(Q), value(Q), color(Q)) (t(j)<t(j+1), 1≤j≤Q-1; Q is the total number of tuples); t(1)=1 and t(Q)=T+1*. The intervals *[t(j), t(j+1)-1] (1≤j≤Q-1)* are colored with the color *color(j)*. After computing these intervals, any two consecutive intervals which have the same color *col* need to be (repeatedly) merged into one larger interval (their union), which has color *col*. The final set of intervals is the set of maximally-colored intervals. The time complexity of this approach is also *O(M·log(M))*. Note that none of the two approaches we presented enforces any limit upon the number of time slots *T* (thus, *T* can be as large as we want). Another possibility is to consider the coloring operations in reverse order. We will use the disjoint sets mechanism (Galil, 1991). Initially, every time slot *t* is alone in a separate set and has *left(t)=right(t)=t*. When we color an interval *[a,b]*, we maintain a counter *idx* which starts at *a* and we will traverse all the yet-uncolored time slots within *[a,b]*. When we reach a time slot *t*, we know if it was previously colored or not. If it wasn't, then we color it and move to the next time slot *t+1*. After coloring a time slot *t*, we immediately check if the time slots *t-1* and *t+1* are also colored (if they exist). If *t* and *t-1* are both colored, we need to combine the sets corresponding to *t* and *t-1*. If *t+1* is also colored, we will then combine the sets of *t* and *t+1*. When combining two sets, we find their two representatives *A* and *B*. A representative *Q* maintains the leftmost and rightmost time slot in the set (because every set is an interval): *left(Q)* and *right(Q)*. We choose *A* or *B* to be the new representative (according to the heuristic we use; e.g. union by rank, or union by size). Let's assume that *A* was chosen as the representative of the combined set. Then we set *left(A)=min{left(A), left(B)}* and *right(A)=max{right(A), right(B)}*. When the counter *idx* reaches a time slot *t* which is already colored, we find the representative *t'* of the set containing *t* and we set *idx=right(t')+1*. This way, every time slot is colored at most once and the time complexity is *O(M+T·log(T))* or (*O(M+T·log\*(T))*) if we also use path compression). Another useful problem in the case of data transfers which require full link usage is to find the longest interval of available time slots. We can support this by using a segment tree or a block partition. With this data structure, we can add the same value to a range of time slots and query the maximum sum segment of slots fully contained inside a given interval *[a,b]*. If we associate to each available time slot *t* a value *v(t)=A>0* and to each occupied time slot a value *v(t)<-T·A*, then the maximum sum segment corresponds to the largest interval of available time slots. A bandwidth reservation is made by adding a value *X<-T·A* to a range of slots *[a,b]* and is cancelled by adding the value *–X* to the same range of slots corresponding to the reservation.

## 4. DATA TRANSFERS IN TREE NETWORKS

A point-to-point data transfer request can specify a minimum required bandwidth or a maximum path delay. Because of this, it is useful to be able to compute efficiently aggregates over values associated to the vertices and edges of a tree network. We consider that every vertex $v$ has a weight $wv(v)$ and every edge $(u,v)$ has a weight $we(u,v)$. We will root the tree at some vertex $r$ (called its root). We are interested in maintaining several types of aggregate information, subject to unexpected edge and vertex weight changes. The kind of information we want to be able to compute efficiently is: 1) what is the aggregate weight of all the edges (vertices) on the path from a root to a given vertex v ? ; 2) what is the aggregate weight of all the edges (vertices) in the subtree of a given vertex $v$ ? ; 3) what is the aggregate weight (e.g. *min, max, +*) of the edges on the path between two vertices *u* and *v* ? First, we will compute a modified Euler tour of the tree. This tour consists of a sequence of *2·n* occurrences of the *n* vertices of the tree. In order to compute the tour, we perform a DFS traversal of the tree starting from the root. We add the vertex *i* at the end of the sequence (initially empty) when we enter vertex *i* from its parent or from the initial call, and when we finish traversing vertex *i*'s subtree (thus, every vertex appears twice, including the leaves). For each vertex *i*, we compute $a(i)$ and $b(i)$, the first and last position on which *i* appears in the Euler tour. We consider aggregation functions *aggf* which have an inverse (e.g. *+, xor*); we denote the inverse of a value *val* by $val^{-1}$, and the neutral value by *e*. For the path aggregate weight case, the weight assigned to a position $a(i)$ is $w(a(i))=we(parent(i), i)$, if $i \neq r$, or *e*, if $i=r$ (or $wv(i)$ in the vertex case), and the weight of a position $b(i)$ is $(we(parent(i),i))^{-1}$, if $i \neq r$, or *e*, if $i=r$ (or $(wv(i))^{-1}$ in the vertex case). Whenever the weight of an edge *(parent(i),i)* (of vertex *i*) changes by *d*, we must change $w(a(i))=aggf(w(a(i)),d)$ and $w(b(i))=aggf(w(b(i)),d^{-1})$. The aggregate weight of the edges (vertices) on a path from the root to a vertex *i* is the aggregate of the weights in the interval *[1,a(i)]*. If we construct a segment tree over the *2·n* positions of the Euler tour (the segment tree has *2·n* leaves), we can compute this value by using a range aggregate query over the corresponding interval in the segment tree. Thus, path aggregate queries and weight (point) updates can be performed in *O(log(n))* time. In order to compute the aggregate weight of the edges (vertices) on a path between two given vertices *u* and *v*, we compute the lowest common ancestor of *u* and *v* (*LCA(u,v)*). Then, we compute the aggregate of the weights on the path from the root to *u*, *v* and *LCA(u,v)* (*aggu, aggv* and *aggLCA*); the result is: $aggf(aggu,aggv,aggLCA^{-1},aggLCA^{-1})$ (in the edge case), or $aggf(aggu,aggv,aggLCA^{-1},aggLCA^{-1},wv(LCA(u,v)))$ (in the vertex case) (see also (Andreica and Tirsa, 2008)). For the second type of queries, we assign weights only to the positions $a(i)$: $w(a(i))=we(parent(i),i)$ (for the edge case), or $wv(i)$ (for the vertex case); $w(b(i))=e$. The aggregate of all the weights in vertex *i*'s subtree is the result of a range query over the interval *[a(i)+1,b(i)]* (for the edge case), or *[a(i),b(i)]* (for the vertex case). Updating the weight of an edge *(parent(i),i)* (vertex *i*) by *d* requires the (point) update of $w(a(i))$ (which must be updated by *d*). Thus, we can use a segment tree in this case, too. For this type of queries, we can replace the value of $b(i)$ by the largest position $a(j) \leq b(i)$. We can do this by maintaining the type (*a* or *b*) and the corresponding vertex *i* of each position *k* in the Euler tour. We traverse the positions from *1* to *2·n*. Whenever we encounter a type *a* position *k*, we set a variable *last_a* to *k* and we increment a counter *cnt_a* by *1* (*cnt_a* is initially *0*). When we encounter a type *b* position *k*, corresponding to a vertex *i*, we set $b(i)$ to *last_a* (or to *cnt_a*, if we later renumber the positions of the tour). Then, we can remove all the positions $b(i)$ from the tour and maintain only *n* values (the $a(i)$ positions, which can now be renumbered from *1* to

$n$). If only the values $a(i)$ and $b(i)$ are given for each vertex $i$ (without the Euler tour itself), we will need to sort these values, in order to obtain the Euler tour first. For the third type of queries, a fully dynamic solution is based on tree decomposition techniques. We will only present a solution for the static case, which is more efficient by an $O(log(n))$ factor than the dynamic case. We will compute the values $Anc(i,j)$=the ancestor of vertex $i$ located $2^j$ levels higher (the level of a vertex $u$ is the distance between $u$ and $r$; $level(r)=0$ and $level(u \neq r)=level(parent(u))+1$), and $Agg(i,j)$=the aggregate of the edge (vertex) weights on the path between $i$ and $Anc(i,j)$. We have $Anc(i \neq r,0)=parent(i)$ ($Anc(r,0)=r$) and $Anc(i,j \geq 1)=Anc(Anc(i,j-1),j-1)$; $Agg(i \neq r,0)=we(parent(i),i)$ for the edge case (or $wv(i)$ for the vertex case) and $Agg(i,j \geq 1)=$if $(level(i) \geq 2^j)$ then $aggf(Agg(i,j-1), Agg(Anc(i,j-1),j-1))$ else undefined. In order to compute the aggregate weight on the path between $u$ and $v$, we first compute $LCA(u,v)$. Then, we will compute the aggregates $aggu$ and $aggv$ on the paths between $u$ and $LCA(u,v)$, and $v$ and $LCA(u,v)$. The answer will be $aggf(aggu, aggv)$ for the edge case (and $aggf(aggu, aggv, wv(LCA(u,v)))$ for the vertex case). In order to compute the aggregate on the path between a vertex $u$ and an ancestor $au$ of $u$, we initialize $j$ to $log(n)$, $pu$ to $u$ and $pagg$ to *undefined*. While $(level(pu)>level(au))$ we perform the following actions: *(1)* as long as $(level(pu)-2^j<level(au))$ we decrease $j$; *(2)* we set $pagg$ to $aggf(pagg, Agg(pu,j))$; *(3)* we set $pu$ to $Anc(pu,j)$. Computing $LCA(u,v)$ is done similarly: we first test if $u$ is an ancestor of $v$ (in which case $LCA(u,v)=u$), or if $v$ is an ancestor of $u$ (in which case $LCA(u,v)=v$); otherwise: *(1)* $j=log(n)$; *(2)* $pu=u$; *(3)* while $(j \geq 0)$ do: { *(3.1)* while $(j \geq 0)$ and $(Anc(pu,j)$ is an ancestor of $v)$ do $j=j-1$; *(3.2)* if $(j \geq 0)$ then $pu=Anc(pu,j)$ }; *(4)* $LCA(u,v)=Anc(pu,0)$. We can test in $O(1)$ time if $a$ is an ancestor of $b$.

## 5. OFFLINE DATA DISTRIBUTION PROBLEMS

### 5.1 Largest Revenue Path with Limited Cost in Trees

We are given a tree with $n$ vertices. Each (undirected) edge $(u,v)$ has a cost $C(u,v)$ and a revenue $P(u,v)$ (both the cost and the revenue are non-negative). For every (unordered) pair of neighboring edges $(u,v)$ and $(u,w)$ we also have a switching cost $SC(u,v,w) \geq 0$ and a switching revenue $SP(u,v,w) \geq 0$. We want to solve a bicriteria data distribution optimization problem. Given an upper limit $C_{max}$, we want to find a path in the tree such that the sum of the costs of the edges on the path (plus the switching costs of any two consecutive edges on the path) is at most $C_{max}$ and the sum of the revenues of the edges on the path (plus the switching revenues of any two consecutive edges on the path) is maximum. We will consider two cases: (1) the degree of every vertex in the tree is bounded by a small constant $D_{max}$; (2) the degrees of the vertices are not bounded, but the switching costs and revenues are all zero. For both cases we will use the same general framework, based on computing the *centroid decomposition* of the given tree. The centroid decomposition of a tree with $n$ vertices in which every vertex $i$ has a positive weight $w(i)$ is defined as follows. First, the centroid of the tree is found. The centroid is a vertex which, if removed, the maximum total weight of the vertices in any connected component of the resulting forest is minimum. A centroid can be computed in linear time for a tree with weighted vertices. We first compute the total weight of the tree, $WTT$. Then, we root the tree at an arbitrary vertex $r$ and we traverse the tree bottom-up (from the leaves towards the root). For each vertex $i$, we compute $WT(i)$=the sum of the weights of the vertices in its subtree $T(i)$. For a leaf vertex $i$, $WT(i)=w(i)$; for a non-leaf vertex $i$, $WT(i)$ is equal to $w(i)$, plus the sum of the values $WT(s(i,j))$ ($1 \leq j \leq ns(i)$) ($ns(i)$=the number of sons of vertex $i$; $s(i,j)$=the $j^{th}$ son of vertex $i$). For each vertex $i$, we also compute $W_{max}(i)=max\{max\{WT(s(i,j))|1 \leq j \leq ns(i)\}, WTT-WT(i)\}$, i.e. the maximum total weight of a connected component, in case vertex $i$ is removed. The centroids are those vertices for which $W_{max}(i)$ is minimum. We choose one of these vertices as the tree centroid. Then, we obtain the connected components, as if the centroid were removed. For each connected component (which is a tree), we compute its centroid decomposition, recursively. We stop when the tree (component) has only one vertex. The centroid decomposition constructs a centroid tree. The centroid $C$ of the tree is the root of the centroid tree. Then, we compute the centroid trees and decompositions (and the centroids $C'$) of the connected components obtained by removing vertex $C$. We make each such centroid $C'$ the son of $C$ (basically, $C$ connects the centroid trees of the components obtained by removing $C$). When $w(i)=1$, the total weight of the vertices of a component is equal to the number of vertices in that component; the height of the centroid tree is $O(log(n))$ in this case (because the number of vertices of each component halves at each step) and the overall complexity of the described algorithm is $O(n \cdot log(n))$. We start by computing the centroid decomposition (centroid tree) of the original tree (with unit vertex weights). As soon as we find the centroid $C$ of a component, we will also compute the best path which passes through that vertex and contains only vertices of that component. To be more precise, at first we compute the best path which passes through the centroid of the initial tree. Any path which does not contain the centroid vertex must be fully contained in one of the components obtained by removing $C$ from the tree. We repeat this procedure recursively for each component. If the time required to compute the best path passing through a given vertex $C$ in a tree with $n$ vertices is $TP(n)$, then the total time required is $O(TP(n)+2 \cdot TP(n/2)+ \ldots +2^i \cdot TP(n/2^i)+ \ldots +n \cdot TP(1))$, which, in the worst case, is $O(TP(n) \cdot log(n))$. We will now explain how to compute the optimal path passing through a specified vertex $r$ of a tree with $n$ vertices (i.e. find $TP(n)$). We will root the tree at the vertex $r$. For each vertex $i$, we will compute $C_{root}(i)$ and $P_{root}(i)$, the total cost and the total revenue of the path from the root $r$ to vertex $i$. $C_{root}(r)=P_{root}(r)=0$, $C_{root}(i \neq r)= C_{root}(parent(i)) + C(parent(i), i) + SC(parent(i), parent(parent(i)), i)$ and $P_{root}(i \neq r) = P_{root}(parent(i))+P(parent(i),i)+SP(parent(i), parent(parent(i)), i)$; if $parent(i)=r$, then $parent(parent(i))$ is not defined and $SC(parent(i), parent(parent(i)), i) = SP(parent(i), parent(parent(i)), i)=0$. A candidate for the optimal path is the path from $r$ to any vertex $i$ with $C_{root}(i) \leq C_{max}$ and a maximum value for $P_{root}(i)$; we denote by $Pr(r)=max\{-\infty$, the revenue of such a candidate path$\}$. We now need to consider paths which start in the subtree of a son $s(r,j_1)$ of the root and end in the subtree of a different son, $s(r,j_2)$ $(j_1 \neq j_2)$. We will first handle case (1). We will traverse all the vertices in the tree. For each vertex $i$, we will maintain the son $pson(i)$ of the root $r$ which is contained on the path from $i$ to $r$ (if $i$ is a son of $r$, then $pson(i)=i$; otherwise, $pson(i)=pson(parent(i)))$. We will maintain a set $S(j)$ for every son $j$ of the root. We will insert each tuple $(i, C_{root}(i), P_{root}(i))$ in $S(pson(i))$. Then, we sort the tuples $(q, C_{root}(q), P_{root}(q))$ in $S(j)$ ($j$ is a son of $r$) in increasing order of their $C_{root}(q)$ values. For a set $S(j)$, let's assume that the order of the tuples is $(q(j,1), C_{root}(q(j,1)), P_{root}(q(j,1))), …, (q(j,|S(j)|), C_{root}(q(j,|S(j)|)), P_{root}(q(j, |S(j)|)))$ (where $|S(j)|$ is the number of tuples in $S(j)$). We will compute $P_{max}(j,i)$, the maximum value of $P_{root}(q(j,k))$, with $1 \leq k \leq i$. We have $P_{max}(j,0)=-\infty$ and $P_{max}(j,1 \leq i \leq |S(j)|)=max\{P_{max}(j,i-1), P_{root}(q(j,i))\}$. Then, we traverse the tree nodes again. For each vertex $i$ with $C_{root}(i) \leq C_{max}$, we will compute the largest revenue path starting at vertex $i$, passing through the root $r$ and ending at another vertex in the tree, such that its total cost is at most $C_{max}$; we denote the revenue of this path by $Pr(i)$ (which is $-\infty$ initially). We will consider every son $j \neq pson(i)$ of the root. For each such son $j$, we will find the largest index $k$ such that $C_{root}(q(j,k)) \leq C_{max}-C_{root}(i)-SC(r, pson(i), j)$. If $k \geq 1$, then we set $Pr(i)=max\{Pr(i), P_{root}(i)+SP(r, pson(i), j)+P_{max}(j,k)\}$. The largest revenue of a path passing through the root $r$ and obeying all the constraints is $max\{Pr(i)|1 \leq i \leq n\}$. In this case, $TP(n)=O(n \cdot log(n)+n \cdot D_{max} \cdot log(n))$ if we sort the tuples for each son using a comparison-based sorting algorithm and we find the largest index $k$ corresponding to a son $j$ of the root (and given a fixed vertex $i$) by binary search, in $O(log(n))$ time. If the cost values are integers and are bounded by a constant $CC_{max}$, then we can sort all the tuples in $O(CC_{max}+n)$ time. Afterwards, we can compute an array $P_{max}'(j)$ for each son $j$ of the root. We initialize all the values in $P_{max}'(j)$ to $-\infty$. Then, we set $P_{max}'(j, C_{root}(q(j,k)))=max\{P_{max}'(j, C_{root}(q(j, k))), P_{root}(q(j,k))\}$. Afterwards, we traverse the entries $cc=1,…,CC_{max}$ and set $P_{max}'(j,cc)=max\{P_{max}'(j,cc-1), P_{max}'(j,cc)\}$. Whenever we want to find the largest revenue of a tuple corresponding to a descendant of a son $j$ of the root, such that the tuple's cost is at most $CG$, we return $P_{max}'(j,CG)$. The time complexity of the algorithm becomes $O(n+CC_{max}+n \cdot D_{max})$. If we consider $CC_{max}$ and $D_{max}$ to be constants, the time complexity is linear ($TP(n)=O(n)$). In order to handle case (2), we can use the same approach as for case (1). However, since the degree of a vertex is not bounded, the time complexity of the proposed solution may become $O(n^2 \cdot log(n))$ (or $O(n^2+CC_{max})$). In order to obtain a better time complexity, we will consider all the tuples $(q, C_{root}(q), P_{root}(q))$ together and sort them in increasing order of $C_{root}(q)$: $(q(1), C_{root}(q(1)), P_{root}(q(1))), …, (q(n), C_{root}(q(n)), P_{root}(q(n)))$. We will compute the values $P_{max}(j)$, defined as follows: $P_{max}(0)=-\infty$, $P_{max}(1 \leq j \leq n)=max\{P_{max}(j-1), P_{root}(q(j))\}$. For each value $P_{max}(j)$ we will store the value $Rson(j)=pson(i)$ where $P_{max}(j)=P_{root}(i)$. $Rson(0)=0$ and $Rson(1 \leq j \leq n)=(if\ P_{max}(j)=P_{max}(j-1)\ then\ Rson(j-1)\ else\ pson(q(j)))$. Afterwards, we will compute the values $P_{max,2}(j)$ and $Rson_2(j)$ ($0 \leq j \leq n$). $P_{max,2}(j)$ is the largest value of $P_{root}(q(k))$ ($1 \leq k \leq j$), such that $pson(q(k)) \neq Rson(j)$. $P_{max,2}(0)=-\infty$ and $Rson_2(0)=0$. For $1 \leq j \leq n$, we consider the three pairs $(pr=P_{max}(j-1), rs=Rson(j-1))$, $(pr=P_{max,2}(j-1), rs=Rson_2(j-1))$, $(pr=P_{root}(q(j)), rs=pson(q(j)))$. We disregard those pairs with $rs=Rson(j)$. From the remaining pairs, we choose the pair $tp$ with the largest value of $tp.pr$ and set $(P_{max,2}(j), Rson_2(j))=(tp.pr, tp.rs)$. Note that there will be at least one remaining pair to choose from. Afterwards, for every vertex $i \neq r$, with $C_{root}(i) \leq C_{max}$, we will compute $Pr(i)$, having the same meaning as for case (1). In order to compute $Pr(i)$, we need to find the optimal path starting at the root, ending at a vertex $j$ with $pson(j) \neq pson(i)$ and whose total cost is at most $C_{limit}=C_{max}-C_{root}(i)$. In order to do this, we find the largest index $k$ such that $C_{root}(q(k)) \leq C_{limit}$. We now consider the two pairs $(pr=P_{max}(k), rs=Rson(k))$, $(pr=P_{max,2}(k), rs=Rson_2(k))$. If one of the pairs, $tp$, has $tp.rs=pson(i)$, then we disregard this pair. Afterwards, we set $Pr(i)=P_{root}(i)+tpmax.pr$, where $tpmax$ is the pair with the largest value of the $pr$ field (among the one or two remaining pairs). $Pr(r)$ is computed just like in case (1). The same holds for computing the largest revenue of a path passing through the root $r$. In this case, $TP(n)=O(n \cdot log(n))$. For the case of bounded integer values of the costs, we can have $TP(n)=O(n+CC_{max})$ (and, if we consider $CC_{max}$ to be a constant, $TP(n)=O(n)$). We can improve the algorithm slightly, if we consider the vertices $i$ in increasing order of their costs $C_{root}(i)$, i.e. in the order $q(1), …, q(n)$. For the first vertex $q(1)$, we start with $k=n$ and decrease $k$ by $1$ until $k=0$ or $C_{root}(q(k)) \leq C_{max}-C_{root}(q(1))$. For $2 \leq i \leq n$, we start with $k$ equal to the index $k$ computed for $q(i-1)$ and continue to decrease it by $1$, until we reach $k=0$ or $C_{root}(q(k)) \leq C_{max}-C_{root}(q(i))$. The time complexity of this stage is $O(n)$, as $k$ is decreased $O(n)$ times. However, we still need to sort the tuples $(q(i), C_{root}(q(i)), P_{root}(q(i)))$ initially.

## 5.2 Offline Data Distribution in Mobile Wireless Path Networks with Immediate Processing Time

We consider a simple model of a wireless path network with $n$ nodes, in which every node $i$ of the network is (initially, at time $0$) located at coordinates $x(i)$ ($1 \leq i \leq n$; $x(i) \leq x(i+1)$). Node $1$ needs to transmit a piece of content to every other node in the network. A node $i$ can transmit the content instantly to a node $j$ if the distance between them is at most $D$ (i.e. $|x(i)-x(j)| \leq D$). Note that a node $j$ can transmit the content further as soon as it receives it. Thus, node $j$ can transmit the content immediately to another node $k$ if $|x(j)-x(k)| \leq D$. Each node is mobile and can travel with (at most) a speed $v$. We want to compute the minimum time duration after which all the nodes receive the content from node $1$. We will present two approaches. The first one is a linear time algorithm. For every node $i$ ($1 \leq i \leq n$) we will compute $Tmin(i)=$the minimum amount of time after which node $i$ can receive the content and $xmax(i)=$the maximum x-coordinate at which node $i$ can be located in order to (still) receive the content by the time moment $Tmin(i)$. $Tmin(n)$ is the minimum time after which all the nodes receive the content. Obviously, we have $Tmin(1)=0$ and $xmax(1)=x(1)$. For $2 \leq i \leq n$ we proceed as follows. If $(x(i)-xmax(i-1)>D)$ then node $i$ needs to get closer to node $i-1$ in order to receive the content. Let $tdif=(x(i)-xmax(i-1)-D)/v$. If $tdif \leq Tmin(i-1)$, then node $i$ travels from time $0$ to time $tdif$ to the coordinate $xmax(i-1)+D$ (at maximal speed) and waits there until node $i-1$ receives the content. When node $i-1$ receives the content, it will immediately send it to node *i*; thus, *Tmin(i)=Tmin(i-1)* and *xmax(i)=xmax(i-1)+D*. If *tdif>Tmin(i-1)*, then node *i* travels from time *0* to time *Tmin(i-1)* to the coordinate *x'(i)=(x(i)-v·Tmin(i-1))* (at maximal speed). At time *Tmin(i-1)*, we have *x'(i)-xmax(i-1)>D*. Let *tdif'=(x'(i)-xmax(i-1)-D)/(2·v)*. From time *Tmin(i-1)* to time *Tmin(i-1)+tdif'*, nodes *i* and *i-1* travel towards each other (at maximal speed). Thus, *Tmin(i)=Tmin(i-1)+tdif'* and *xmax(i)=x(i)-Tmin(i)·v*. If the initial distance *x(i)-xmax(i-1)* is at most *D*, then node *i* will move away from node *i-1*. Let *tdif=(xmax(i-1)+D-x(i))/v*. Node *i* travels from time *0* to time *min{Tmin(i-1), tdif}* to coordinate *xmax(i)=x(i) + v·min{Tmin(i-1), tdif}* (at maximal speed) and then waits there until time *Tmin(i-1)*. We have *Tmin(i)=Tmin(i-1)*. The second approach is based on binary searching the minimum value *Tmin* after which all the nodes receive the content. The feasibility test consists of computing *xmin(i)*=the minimum x-coordinate at which node *i* can be located at the moment of receiving the content, such that node *n* can still receive the content by the time moment *Tmin*, and *Tmax(i)*=the largest time moment at which node *i* can receive the content, such that node *n* can still receive the content by time *Tmin*. If the value is feasible then we will test a smaller value next; otherwise, we will test a larger value next. We will describe the feasibility test next. *Tmax(n)=Tmin* and *xmin(n)=x(n)-v·Tmin*. For *1≤i≤n-1* (in decreasing order), we proceed as follows. If *xmin(i+1)-x(i)>D*, then let *tdif=(xmin(i+1)-D-x(i))/v*. We have *Tmax(i)=Tmax(i+1)-tdif* and *xmin(i)=x(i)*. If *x(i)≥xmin(i+1)-D*, then let *tdif=min{(x(i)-xmin(i+1)+D)/v, Tmax(i+1)}*. Node *i* travels from time *0* to time *tdif* to coordinate *xmin(i)=x(i)-tdif·v* and then waits there. We have *Tmax(i)=Tmax(i+1)*. If, at some point, *Tmax(i)* drops below *0*, or *(x(i)+Tmax(i)·v<xmin(i)* for some node *i*), then *Tmin* is not a feasible value. The time complexity of this approach is *O(n·log(TM))*, where *TM* is a good upper bound for the time duration we were searching for.

### 5.3 Offline Data Distribution in Wireless (Path) Sensor Networks with Release Times

In this subsection we consider a problem which is similar to the one from the previous subsection. *n* wireless network nodes are located on the real line (node *i* is located at position *x(i)*; *1≤i≤n*), such that *x(1)≤x(2)≤…≤x(n)*. Node *1* has a piece of content which it needs to distribute to all the other nodes. The nodes are very simple processing devices (e.g. sensor nodes) and every node *i* is connected only to the nodes immediately to its left and to its right (*i-1* and *i+1*, if they exist). If a node *2≤i<n* receives the content at a time *t≥0*, it performs the following actions: if it did not receive the content before and *t<pt(i)*, it can wait until the time moment *pt(i)* (if it so wishes); let's denote *t'=t* (if it chooses not to wait) or *t'=pt(i)* (if it chooses to wait); if *t'≥pt(i)*, then it processes the content, which takes a duration *d(i)*. Afterwards, if the content was received from its left (right) neighbour, it forwards it to its right (left) neighbour. When node *n* receives the content at time *t*, if *t<pt(i)*, then it waits until *t=pt(i)*; afterwards, it processes the content (which takes a duration *d(n)*) and then sends it back to node *n-1*. The time values *pt(i)* are the processing release times for each node *i*. Node *i* cannot start processing before the time moment *pt(i)* (considering that the initial time moment is *0*), due to several factors (e.g. in order to save energy, it can only perform processing tasks during certain time periods). The content travels at a speed *s*; thus, the duration of sending the content from a node *i* to a neighbouring node *j* is *|x(i)-x(j)|/s*. We consider here only the restricted case where *pt(i)≤pt(i+1)* (*2≤i≤n-1*). For this case, when node *1* receives the content back, it knows that all the other nodes have received the content (it is easy to prove that this is the case). It is also easy to notice that the content is first sent from left to right (the *left-to-right pass*) and then it is sent back, from right to left (the *right-to-left pass*). We want to minimize the time duration after which node *1* receives the content back (which acts as an acknowledgement). The duration is influenced by the local *waiting* decisions made by each node. The considered problem is offline, because we will globally make these decisions and the problem parameters are fixed. We will first consider the case where *d(i)=0* (*2≤i≤n-1*). In this case, no node *i* (*2≤i≤n-1*) chooses to wait (if it has the opportunity). The content reaches node *n* at time *t=|x(n)-x(1)|/s*. If *t<pt(n)*, then node *n* waits until time *pt(n)*. Afterwards, it sends the content back. This time, because *pt(i)≤pt(i+1)*, every node which did not process the message during the left-to-right pass, will process the message now. The total duration is *max{|x(n)-x(1)|/s, pt(n)}+d(n)+|x(n)-x(1)|/s*. We will now consider a second easier case, in which all the values *d(i)* (*2≤i≤n-1*) are equal (thus, we will say that *d(i)=dp*). We will use a dynamic programming algorithm and compute the values *Twmin(i,j)*=the minimum total waiting time during the left-to-right pass if the content reached node *i* and *j<i* nodes processed the content so far. We have *Twmin(1≤i≤n-1, 0)=0* and *Twmin(i,j≥i)=+∞*. In order to compute *Twmin(i, 1≤j≤i-1)*, we consider the values *Twmin(i-1,j-1)* and *Twmin(i-1,j)*. We first initialize *Twmin(i,j)=+∞*. For the case *Twmin(i-1,j-1)*, we compute the time moment when the content reaches node *i*, which is *tr=|x(i)-x(1)|/s+Twmin(i-1,j-1)+(j-1)·dp*. If *tr≤pt(i)*, then *Twmin(i,j)=min{Twmin(i,j), Twmin(i-1,j-1)+ (pt(i)-tr)}*; else, *Twmin(i,j)=min{Twmin(i,j),Twmin(i-1,j-1)}*. We then compute *tr2=|x(i)-x(1)|/s+Twmin(i-1,j)+j·dp*. If *tr2<pt(i)* then *Twmin(i,j)=min{Twmin(i,j), Twmin(i-1,j)}* (node *i* chooses not to wait). After computing all these values, we will compute the minimum total waiting time *Tmin* (initially set to *+∞*), based on the values *Twmin(n-1,\*)* and the decisions made by node *n*. We will consider all the values *j* (*0≤j≤n-2*). For each value, we compute the time moment *tr=|x(n)-x(1)|/s+Twmin(n-1,j)+j·dp* when the content reaches node *n*. If *tr<pt(n)*, we set *Tmin=min{Tmin, Twmin(n-1,j)+pt(n)-tr}*; otherwise, we set *Tmin=min{Tmin, Twmin(n-1,j)}*. The total duration (before the content returns to node *1*) is equal to *2·|x(n)-x(1)|/s+(n-2)·dp+d(n)+Tmin*. As we can see, the only term which can be minimized is *Tmin* (the others are independent of the chosen distribution strategy). The time complexity is *O(n²)*. For the general case, where the *d(\*)* values may be different, we present a pseudo-polynomial solution when the durations *d(\*)* are integers. We compute *Twmin(i, tproc)*=the minimum total waiting time during the left-to-right pass if, by the time the message leaves node *i*, *tproc* time units were spent by all the nodes (so far) with the content processing. We have *Twmin(1,0)=0* and *Twmin(i,*

$tproc>sd(i))=+\infty$ (where $sd(i)=d(1)+...+d(i)$; $d(1)=0$). In order to compute the $Twmin(i\geq 2,*)$ values, we will first initialize them to $+\infty$. Afterwards, we consider all the values $Twmin(i-1, tproc)$. For each pair $(i-1, tproc)$, we compute $tr=|x(i)-x(1)|/s+Twmin(i-1,tproc)+tproc$. If $tr<pt(i)$, then we set $Twmin(i, tproc)=min\{Twmin(i, tproc), Twmin(i-1, tproc)\}$ (node $i$ chooses not to wait) and $Twmin(i, tproc+d(i))=min\{Twmin(i, tproc+d(i)), Twmin(i-1, tproc)+pt(i)-tr\}$ (node $i$ chooses to wait); if $tr\geq pt(i)$, we set $Twmin(i, tproc+d(i))=min\{Twmin(i, tproc+d(i)), Twmin(i-1, tproc)\}$. After this stage, we will compute the same value $Tmin$ as before. For every pair $(n-1, tproc)$ we compute $tr$, the time moment when the content reaches node $n$ ($tr=|x(n)-x(1)|/s+Twmin(n-1, tproc)+tproc$) and if $tr<pt(n)$, we set $Tmin=min\{Tmin, Twmin(n-1, tproc)+pt(n)-tr\}$; otherwise, $Tmin=min\{Tmin, Twmin(n-1, tproc)\}$. The total duration will be equal to $2\cdot |x(n)-x(1)|/s+Tmin+(d(2)+...+d(n))$. The time complexity is $O(n\cdot TMAX)$ ($TMAX=d(2)+...+d(n)$).

*5.4 Packet Permutations with k increasing 2-sequences*

We consider a communication flow composed of $n$ packets (numbered from *1* to *n*). Each packet $i$ contains checksum information about packet $i-1$. If the packets are sent in the normal logical order, we want to know how many possible receiving orders exist in which (exactly) $k$ pairs of packets $(i, i+1)$ arrive immediately one after another. This is the same as computing the number of $n$-element permutations with $k$ increasing 2-sequences. We will compute the values $P(i,k)$=the number of $i$-element permutations with $k$ increasing 2-sequences ($0\leq k\leq i-1$). We will consider $P(i,k)=0$ for $k<0$ or $k\geq i$. We have $P(1,0)=1$. For $i>1$ and $0\leq k\leq i-1$, we have $P(i,k) = (i-k-1)\cdot P(i-1,k) + (k+1)\cdot P(i-1,k+1) + P(i-1,k-1)$. The *3* terms correspond to the following situations: 1) there are $(i-k-1)$ positions where element $i$ can be inserted into an $(i-1)$-element permutation having $k$ 2-sequences, without modifying the number of 2-sequences; 2) there are $(k+1)$ positions where element $i$ can be inserted into an $(i-1)$-element permutation having $k+1$ 2-sequences, in order to "break" one 2-sequence (thus obtaining $k$ 2-sequences); 3) there is one position where we can insert element $i$ into an $(i-1)$-element permutation having $k-1$ 2-sequences, in order to form a new 2-sequence (we insert it right after element $i-1$). The time complexity is $O(n^2\cdot Op(n))$, where $Op(n)$ is the complexity of performing arithmetic operations on the numbers $P(*,*)$ (if the numbers have $O(n)$ digits, then $Op(n)=O(n)$; if we perform all the operations *modulo* a small number $M$, then $Op(n)=O(1)$). $P(n,k)$ is also the number of $n$-element permutations and $k$ decreasing 2-sequences, as the bijective function $f(i)=n-i+1$ maps a permutation with $k$ increasing 2-sequences to one with $k$ decreasing 2-sequences.

## 6. RELATED WORK

In (Henzinger et al., 2003), efficient algorithms are presented for offline and online scheduling of unit capacity multicast data transfers in trees and meshes. In (Andreica and Tapus, 2008), the authors present an algorithmic framework for several efficient data structures which can be used for data transfer scheduling on single-link and path networks. In (Andreica and Tirsa, 2008), the authors present a range of algorithmic techniques for scheduling data transfers in networks with tree topologies. Several heuristic data request scheduling methods were presented in (Theys et al., 2001). A framework for reliable and efficient data placement in distributed systems was presented in (Kosar and Livny, 2005). A scheduling model using bandwidth reservations for critical data transfers was presented in (Hangan et al, 2007).

## 7. CONCLUSIONS AND FUTURE WORK

In this paper we introduced the architecture of a centralized scheduling framework for data transfers in distributed systems. We also took the first steps towards developing efficient algorithmic techniques for scheduling data transfers in distributed systems with arbitrary topologies, by presenting novel methods for handling preemptive and non-preemptive data transfer requests on single network links and in trees. Moreover, we considered several offline data distribution problems, for which we developed new algorithmic solutions.